\documentclass[twocolumn,showpacs,preprintnumbers,amsmath,amssymb]{revtex4}

\usepackage{graphicx}% Include figure files
\usepackage{dcolumn}% Align table columns on decimal point
\usepackage{bm}% bold math

\begin{document}

%\preprint{APS/123-QED}

\title{\large Multiple transitions of the spin configuration in quantum
dots}

\author{M.~C. Rogge}
\email{rogge@nano.uni-hannover.de}
\author{C. F\"uhner}
\author{R.~J. Haug}
 \affiliation{Institut f\"ur Festk\"orperphysik, Universit\"at Hannover,
  Appelstra\ss{}e 2, D-30167 Hannover, Germany}

\date{\today}

\begin{abstract}
Single electron tunneling is studied in a many electron quantum
dot in high magnetic fields. For such a system multiple
transitions of the spin configuration are theoretically predicted.
With a combination of spin blockade and Kondo effect we are able
to detect five regions with different spin configurations.
Transitions are induced with changing electron numbers.
\end{abstract}

% PACS, the Physics and Astronomy   73.63.Kv QDs; electronic transport 73.23.Hk Coulomb blockade; single-electron tunneling 73.21.La QDs; electron states
\maketitle

%%%%%%%%%%%%%%%%%%%%%%%%%%%%%%%%%%%%%%%%%%%%%%%%%%%%%%%%%%%%%%%%%%%%%
%%%%%  Einleitung Physik  %%%%%%%%%%%%%%%%%%%%%%%%%%%%%%%%%%%%%%%%%%%
%%%%%%%%%%%%%%%%%%%%%%%%%%%%%%%%%%%%%%%%%%%%%%%%%%%%%%%%%%%%%%%%%%%%%

%\section{Introduction}

Spin physics in semiconductor quantum dots \cite{Kouwenhoven-97}
has pushed the research on nanostructures, since quantum dots were
suggested as essential elements in quantum computing devices
\cite{Loss-98}. The spin of a single electron on a quantum dot can
act as a qubit and thus great efforts were made to prepare,
manipulate and read the spin on few electron quantum dots. But
also the research on many electron dots has come up with exciting
results and new effects dealing with spin. While it is possible to
detect the spin orientation on very few electron dots just by
identifying energy levels, conclusions about the spin
configuration in many electron dots must be drawn indirectly from
additional spin dependent effects. A few years ago it was
demonstrated that so called spin blockade can be used to read the
spin of the tunnelling electron in lateral quantum dots in a
perpendicular magnetic field \cite{Ciorga-00}. More information
about the spin configuration can be obtained when a Kondo effect
\cite{Kondo-64} is observed leading to spin dependent conductance
in Coulomb blocked regions \cite{Glazman-88,Ng-88}.

\begin{figure}
\centering
\includegraphics[scale=0.8]{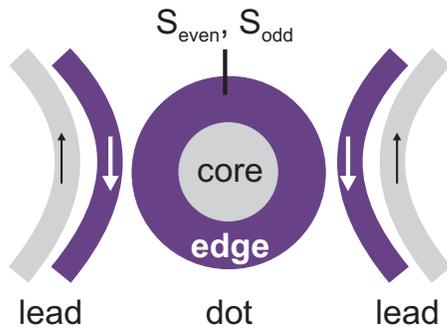}
\caption{Schematic diagram for a lateral quantum dot in a high
perpendicular magnetic field. A shell structure with two Landau
levels is established in the dot, level 0 at the edge (with total
spin $S_{even}, S_{odd}$) and level 1 in the core. The leads shown
are spin polarized due to the spatial separation of edge
channels.} \label{fig1}
\end{figure}

In a perpendicular magnetic field with two Landau levels in the
quantum dot (filling factor $4>\nu_{dot}>2$) a shell structure is
assumed for the dot with Landau level 0 in the edge and level 1 in
the core (see Fig. \ref{fig1}) \cite{McEuen-92}. An unpolarized
spin configuration is expected at the edge of the dot for low
electron numbers in high magnetic fields \cite{Stopa-03}. That
means the total spin at the edge of the dot is 0 whenever the
electron number is even, and $\frac{1}{2}$ when the electron
number is odd: $(S_{even},S_{odd})=(0,\frac{1}{2})$ with
$S_{even}$ and $S_{odd}$ the total spin at the edge of the dot for
even and odd electron numbers (see Fig. \ref{fig1}). A few years
ago multiple transitions of the spin configuration were
theoretically predicted and calculated at the $\nu_{dot}=2$
boundary \cite{Wensauer-03}. The spin polarization should increase
with increasing electron number, $(0,\frac{1}{2}) \rightarrow
(1,\frac{1}{2}) \rightarrow (1,\frac{3}{2})$.... These transitions
of the spin configuration are related to spin flips in the
$\nu_{dot}<2$ regime when the spin configuration approaches the
maximum density droplet. The first two spin configurations
$(0,\frac{1}{2})$ and $(1,\frac{1}{2})$ can be identified with the
combination of spin blockade and Kondo effect \cite{Kupidura-06}.
While the first transition between these two configurations was
observed in a few electron quantum dot \cite{Ciorga-02},
experimental evidence for multiple transitions was never found.

We close this gap and present measurements of multiple transitions
in the magnetoconductance of a many electron quantum dot. We make
use of the combination of both spin blockade and Kondo effect at
$4>\nu_{dot}>2$ to identify five regions with different spin
configurations. Transitions between these regions are induced with
changing electron number as predicted. Nevertheless some
observations imply slight deviations from the theoretical picture.
Evidence is found for a change of spin configuration back and
forth (e.g. $(0,\frac{1}{2}) \rightarrow (1,\frac{1}{2})
\rightarrow (0,\frac{1}{2})$) instead of a continuously increasing
spin polarization. These results as well as results about the
position of the transitions as a function of magnetic field should
affect eventually forthcoming calculations for $\nu_{dot}>2$.

\begin{figure}
\centering
\includegraphics[scale=1]{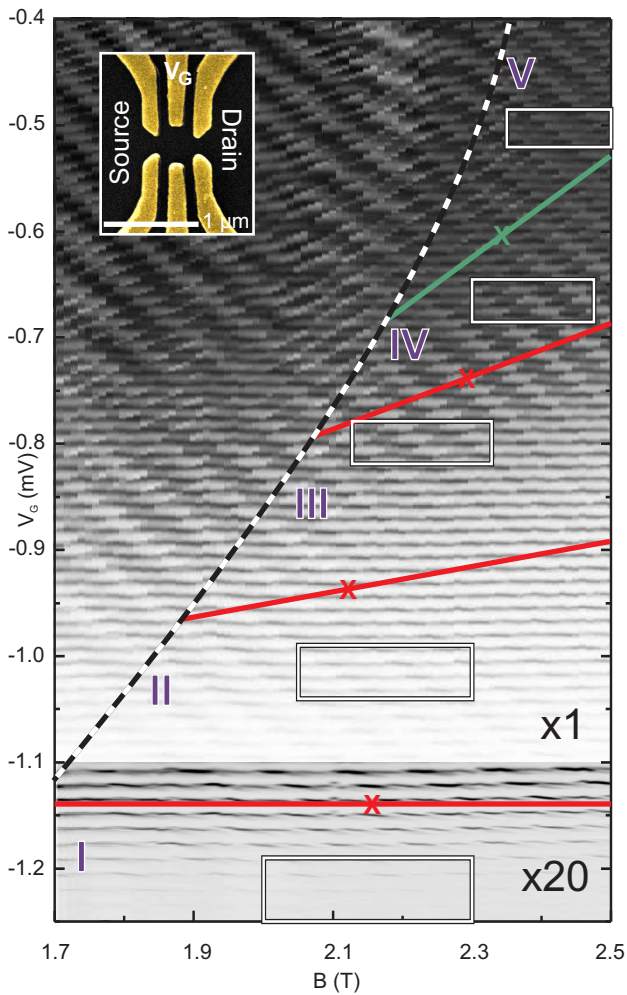}
\caption{Differential conductance $G$ as a function of $B$ and
$V_G$ (dark for high $G$, white for low $G$). On the right of the
dashed line a regular chessboard pattern is observed due to Kondo
effect and spin blockade. The combination of both effects yields
five different regions. The transitions between these regions are
marked with solid lines. They are studied in detail later at the
positions marked with crosses. Inset: SEM picture of our device.}
\label{fig2}
\end{figure}

Our sample \cite{Fuhner-02} is fabricated in split-gate-technology
on a GaAs/AlGaAs heterostructure with a two dimensional electron
system (2DES) 57~nm below the surface (see inset of Fig.
\ref{fig2}). The electron density is $n=3.7*10^{15}$~m$^{-2}$, the
mobility is $\mu=130$ m$^2$/Vs. An electronic dot diameter of
approx. 250~nm yields an electron number of approx. 180.
Differential conductance measurements were carried out in a
$^3$He-$^4$He dilution refrigerator at a base temperature of 20~mK
using standard lock-in technique. With the gate voltage $V_G$ the
dot was tuned in a Kondo regime where signatures of finite Kondo
conductance were studied in detail as a function of a magnetic
field $B$ applied perpendicular to the sample surface.

Such a measurement is shown in Fig. \ref{fig2}. The differential
conductance $G$ is measured as a function of $B$ and $V_G$. Many
Coulomb peaks are visible reflecting changes of the electron
number by one for each peak (nearly horizontal lines). The
depicted range of the gate voltage shows 75 peaks corresponding to
a change of the electron number by 75 electrons. In the Coulomb
blockade regions between the peaks we find a finite conductance
due to Kondo effect which appears because of second order
cotunneling processes. This leads to a chessboard like pattern
\cite{Keller-01,Stopa-03} of high and low differential
conductance. We concentrate our analysis on the right side of the
dashed line were the pattern is most regular. Here in the
$4>\nu_{dot}>2$ regime (the filling factor of the 2DES
$\nu_{2DES}$ is well above 6 in the magnetic field studied here)
only the first Landau level situated at the edge of the dot is
involved in the Kondo conductance. Thus the Kondo effect depends
on the spin configuration at the edge which is different for even
and for odd electron numbers ($S_{even}\neq S_{odd}$). Therefore
the Kondo effect depends on the parity of the electron number at
the edge of the dot (just parity in the following) which is
changed either by electron exchange with the leads induced by gate
voltage or by electron exchange with the center of the dot induced
by the magnetic field.

Apart from the alternating Kondo conductance there is also an
alternating peak amplitude of the Coulomb peaks with the same
periodicity as the Kondo pattern. This is due to spin blockade
\cite{Ciorga-00}. The magnetic field separates the spin polarized
edge channels in the leads (see Fig \ref{fig1}) and thus the
tunneling probability and as a consequence the differential
conductance becomes spin dependent. Spin up transport is
suppressed compared to spin down transport. Spin blockade and
Kondo effect can be observed simultaneously in an intermediate
field regime \cite{Kupidura-06}. Since the periodicity of both
effects is identical the combination establishes a stable pattern
which should not change, if the spin configurations for even and
odd electron numbers at the edge of the dot do not change. As a
consequence changes of this pattern should reflect changes in the
spin configuration at the edge of the dot. If the spin
configuration has changed for one parity (e.g. $(0,\frac{1}{2})
\rightarrow (1,\frac{1}{2})$) the pattern is swaped
\cite{Kupidura-06}.

\begin{figure}
\centering
\includegraphics[scale=1]{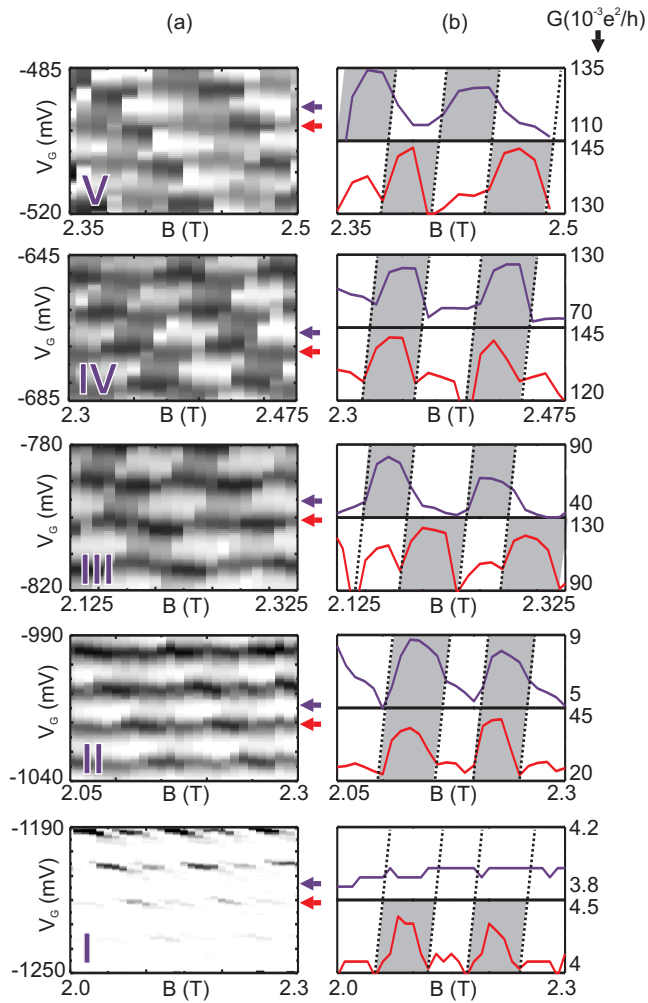}
\caption{(a) enhanced plots of the insets marked in Fig.
\ref{fig2}. In all five regions amplitude modulations of the
Coulomb peaks appear due to spin blockade. Apart from region I a
clear chessboard pattern is visible due to Kondo effect as well.
The combination of both effects changes from region to region. In
II and IV the Kondo effect goes along with a strong amplitude of
the Coulomb peak below while the pattern is swaped in III and V.
(b) The combination of both effects comes out clearer with the
amplitudes of peaks and valleys plotted as a function of $B$. For
each region the lower graph shows the spin blockade, the upper
graph shows the Kondo effect.} \label{fig3}
\end{figure}

In our sample we find not less than 5 regions with different
combinations of Kondo effect and spin blockade. These are marked
with roman numbers I to V in Fig. \ref{fig2}. Therefore we have 4
transitions in between marked with solid lines. Insets in these
regions marked with white boxes in Fig. \ref{fig2} are shown in
detail in Fig. \ref{fig3}.

Fig. \ref{fig3}(a) shows the differential conductance again as a
function of $V_G$ and $B$ for all 5 regions. In Fig. \ref{fig3}(b)
traces are shown reflecting the amplitude along a Coulomb peak
influenced by spin blockade and along the Coulomb valley above
showing Kondo conductance.

In region I no Kondo conductance and thus no chessboard pattern is
observed since the barriers connecting the dot to the leads are
too opaque to allow for second order cotunneling. Nevertheless a
clear bimodal pattern of spin blockade is visible.

In region II Kondo conductance sets in (probably not spin
$\frac{1}{2}$ Kondo \cite{Fuhner-02}) and we can observe a clear
chessboard pattern. We find that finite Kondo conductance appears
above an unsuppressed peak amplitude. Therefore the Kondo effect
is connected to a spin down electron added to the edge of the dot.

This pattern is swaped in region III. Here we find the Kondo
effect above the suppressed peak amplitude. Thus the Kondo effect
is induced by a spin up electron.

While region IV corresponds to region II, region V shows again the
features of region III.

Now that we have identified these 5 regions we will focus on the
details of the transitions between them. This is done exemplarily
at the transition from region III to region IV.

\begin{figure}
\centering
\includegraphics[scale=1]{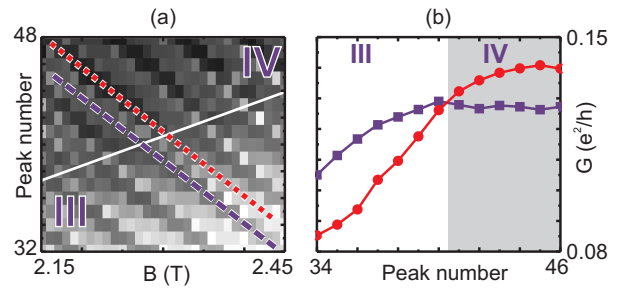}
\caption{(a) peak amplitude as a function of $B$ and peak number
at the transition from III to IV showing only features of spin
blockade. What is a bimodal pattern in Fig.\ref{fig2} now comes
out as a pattern of lines with negative slope. At the transition
the amplitude of these lines swaps reflecting a change of the spin
configuration. Strong lines become weak and vice versa. (b) line
amplitude along the lines marked dashed and dotted in (a). A
crossing appears between peaks 40 and 41.} \label{fig4}
\end{figure}

In Fig. \ref{fig4}(a) we have plotted the peak amplitude in grey
scale as a function of $B$ and the peak number. The resulting plot
is comparable to the original measurement without the Coulomb
valleys. What is an alternating behaviour of peak amplitudes now
becomes a pattern of lines with negative slope. A change of this
pattern should directly reflect a change in the way electronic
spins enter the dot. This in turn corresponds to a change of the
spin configuration in the edge of the dot which is exactly what is
observed in Fig. \ref{fig4}(a). Lines (dashed) with a strong
amplitude below the transition from III to IV (which is marked
with a solid line) corresponding to spin down transport become
weak above the transition now showing spin up transport. In
analogy those lines (dotted) with a weak amplitude below the
transition become strong above the transition. This is shown in
more detail in Fig. \ref{fig4}(b). Traces along lines marked
dashed and dotted in (a) are plotted as a function of the peak
number. A clear crossing of the peak amplitudes is visible between
peak 40 and 41. Thus the origin for the swap of the total pattern
at the transition from region III to IV is identified with the
spin blockade signal. The spin filling mechanism has swaped and
the spin configuration has changed for one parity. The spin
configuration underlying the Kondo effect is not changed as the
Kondo effect is not influenced by this transition. Thus the change
in spin polarization must have happened for the parity, where no
Kondo effect appeared.

\begin{figure}
\centering
\includegraphics[scale=1]{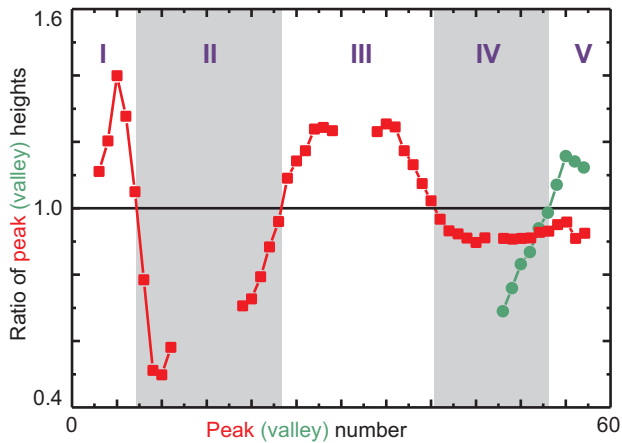}
\caption{For transitions I to IV the ratio of the amplitude of
lines similar to those in Fig. \ref{fig4}(b) is plotted as a
function of the peak number (squares). Swaps corresponding to
changes of the spin configuration appear at the transitions I to
II, II to III and III to IV. No swap was observed at the last
transition. Here the Kondo pattern swaps (circles). Please note
that the data were not taken at the same magnetic field but at the
points marked with crosses in Fig. \ref{fig2}.} \label{fig5}
\end{figure}

Similar analysis were made at all other transitions. The results
are shown in Fig. \ref{fig5}. Again two neighboring lines were
investigated for each transition as shown before in Fig.
\ref{fig4}. But this time for each transition the ratio of the
amplitude of these lines is plotted as a function of the peak
number. If there was no spin blockade, one would expect both peaks
to be equal and thus the ratio would be 1. We find a swap in the
peak ratio at the transitions I to II, II to III and III to IV
which indicates a change in the spin configuration at these
transitions. At the first transition no Kondo effect is visible.
At the other two transitions no significant influence is found on
the Kondo chessboard pattern. As discussed before this implies
that the spin configuration for the parity underlying the Kondo
effect is not changed although two swaps of the spin blockade and
thus two changes of the spin polarization have occurred (one at II
to III, another at III to IV). In our opinion this is not possible
if the spin polarization would have changed twice in the same
direction (e.g. $(0,\frac{1}{2}) \rightarrow (1,\frac{1}{2})
\rightarrow (1,\frac{3}{2})$). If so the spin configuration would
have changed for both parities which should have affected the
Kondo effect. Instead we believe to observe a switching of the
polarization back and forth (e.g. $(0,\frac{1}{2}) \rightarrow
(1,\frac{1}{2}) \rightarrow (0,\frac{1}{2})$). Thus only for one
parity (the one without Kondo effect) the configuration is
changed.

The last transition from region IV to region V is different. Here
the spin blockade stays unchanged, instead we find a swap of the
Kondo pattern marked with circles in Fig. \ref{fig5}. We believe
that this is due to a sudden change of the nature of the Kondo
effect. This might have happened without changing the spin
configurations, as no swap in the spin blockade is observed.
However the slope of the transition line from region IV to region
V marked with a solid line in Fig. \ref{fig2} implies a different
explanation. The slope seems to fit very nicely the slopes of all
other transitions. Thus one would expect the same physical origin
to be responsible. This could be possible if two swaps at once are
assumed changing the spin configuration for both parities (e.g.
$(0,\frac{1}{2}) \rightarrow (1,\frac{3}{2})$). In this case the
spin blockade is not affected and the change in the nature of the
Kondo effect is just due to a new spin configuration.

In conclusion we have found multiple transitions of the spin
configuration in a lateral quantum dot in high magnetic fields as
predicted theoretically. These transitions are detected as changes
of a combined spin blockade/Kondo effect chessboard pattern. They
appear either as a swap of the spin blockade pattern or as a swap
in the Kondo pattern. Regardless the different form of appearance
the slopes of all transitions yield a common origin. They reflect
a change of the spin configuration at the edge of the
two-Landau-level quantum dot with increasing electron number. In
addition the combination of spin blockade and Kondo effect
indicates that a change of the spin polarization back and forth is
observed rather than a continuously increasing spin polarization.

This work has been supported by BMBF.

%%%%%%%%%%%%%%%%%%%%%%%%%%%%%%%%%%%%%%%%%%%%%%%%%%%%%%%%%%%%%%%%%%%%%
%%%%%  Referenzen  %%%%%%%%%%%%%%%%%%%%%%%%%%%%%%%%%%%%%%%%%%%%%%%%%%
%%%%%%%%%%%%%%%%%%%%%%%%%%%%%%%%%%%%%%%%%%%%%%%%%%%%%%%%%%%%%%%%%%%%%

%\newpage

%\bibliographystyle{../../paper/bibtex/apsrev}
%\bibliography{../../paper/bibtex/technology,../../paper/bibtex/dot,../../paper/bibtex/dodo,../../paper/bibtex/qc}

%GATHER{../../paper/bibtex/technology.bib}
%GATHER{../../paper/bibtex/dot.bib}
%GATHER{../../paper/bibtex/dodo.bib}
%GATHER{../../paper/bibtex/qc.bib}

% from dodo.bbl

\end{document}